# The optical generation and continuous transformation of plasmonic skyrmions

ZHE SHEN*, SEN LU, AND XIONGXIONG

*School of Electronic and Optical Engineering, Nanjing University of Science and Technology, Nanjing 210094, China*

**shenzhe@njust.edu.cn*

**Abstract:** Topological quasiparticles, including skyrmions and merons, are topological textures with sophisticated vectorial structures that can be used for high-density information storage, precision metrology, position sensing, etc. Here, we realized the optical generation and continuous transformation of plasmonic field skyrmions. We generated the isolated Néel-type skyrmion using surface plasmon polaritons (SPPs) excited by a focused structured light on a silver film. We used a square and a hexagonal aperture for symmetry constraints and successfully generated the meron lattice and the skyrmion lattice. We unveiled the mechanism of topological texture generation and transformation and optimized the distribution of skyrmion and meron topologies. We further demonstrated the continuous transformation among the isolated skyrmion, the meron lattice, and the skyrmion lattice using well-designed circular-fourfold, circular-sixfold, and fourfold-sixfold symmetry apertures, respectively. This work can open up a pathway for the generation and transformation of skyrmion and meron topologies, which is expected to facilitate new applications in optical information storage and encoding.

## 1. Introduction

In 1962, the British physicist Tony Skyrme found a kind of quasiparticle with topological protection by solving nonlinear field theory equations, namely skyrmion [1]. Possessing a topological non-trivial spin texture, it is the product of the combination of topology and condensed matter, demonstrated in liquid crystals [2], Bose-Einstein condensates [3], chiral magnets [4], and so on. Similar to skyrmions, merons, also known as half-skyrmions, are topological quasiparticles with half-integer topological charges, which have been discovered in the form of pairs or square lattices [5,6]. In past decades, skyrmion and meron topologies have been extensively studied in the field of magnetic materials due to their small size, low current drive, and topological stability [7], and are expected to function as the next-generation carrier for ultra-high density information storage. As the counterpart of magnetic skyrmion and meron topologies, the topological quasiparticles relying on optical methods also received wide attention and showed great advantages in dynamic information transmission, high-resolution imaging, and precision metrology. The latest research realized the propagation-invariant skyrmionic texture in Bessel profiles, which is significant for the skyrmionic beam [8].

The optical skyrmions were first observed in the evanescent electric field of surface plasmon polaritons (SPPs) generated on the hexagonal metal grating structure [9]. Following this configuration, many studies have been reported about the physical properties [10-12] and the dynamic control [13] of the electric field skyrmions. These excellent studies unveiled the fascinating properties of skyrmions and enriched the light-matter interactions at the sub-wavelength scale. However, these skyrmion generation methods strongly depend on the hexagonal metal grating structure or the use of multiple beams, which are complex and undoubtedly increase the complexity and cost of the optical system. Notably, as an ideal physical system supporting skyrmions, SPPs can be excited not only by grating coupling [14] but also by focused light [15-19]. The focused method for SPP excitation makes it possible for the all-optical generation of plasmonic skyrmions, i.e., no need to fabricate a grating structure.

The most representative example is photonic skyrmions [20], whose topological textures rely on the spin angular momentum of the plasmonic field. The generation and manipulation of photonic skyrmions attract much attention and have been further studied [21-24]. However, the field skyrmion, i.e., topological textures rely on the electric field vectors, based on the same focused method is rarely reported. Known as dynamic skyrmions [25], it is worth studying.

In the field of magnetic materials, the transformation [26-31] between skyrmion and meron topologies has been a hot topic because of its significant potential in advanced information-processing devices and information encoding. Correspondingly, the topological transformation in optics is equivalently promising. Recently, a topological transformation work in optics was reported for the first time [32], in which a metallic metasurface was designed and the spin textures were tuned with wavelengths. However, this transformation is discrete, which cannot visualize the continuous evolution between different topologies. Studying and revealing the continuous evolution between different topologies is important for understanding the intrinsic connection between topological quasiparticles. Therefore, it is highly desirable to realize the topological transformation in optics based on the dynamic electric field.

In this study, we realized the optical generation and continuous transformation of plasmonic field skyrmions. A focused method was used to excite SPPs, and the Richard-Wolf theory was used to derive the analytical solution of the plasmonic field. Benefiting from the structureless excitation of skyrmions in a dynamic configuration, this method is convenient and does not depend on the complex hexagonal metal grating structure. Since field symmetry is the key to determining the topology [21], we employed a square and a hexagonal aperture for symmetry constraints to generate the meron lattice and the skyrmion lattice. We further illustrated the factors that affect the distribution of skyrmion and meron topologies and provided an optimized solution. On this basis, we designed three double symmetry apertures and demonstrated the continuous transformation between skyrmion and meron topologies. This study gives a simple method for the generation and transformation of skyrmions and provides a deep understanding of topology, as well as showing significant potential in optical information storage and encoding.

## 2. Theory and method

### *2.1 The topology number of optical skyrmions*

The topological properties of optical skyrmions are characterized by the skyrmion number, which can be expressed as [9,33]:

$$N = \frac{1}{4\pi}\iint_{\sigma} \vec{e} \cdot \left( \frac{\partial \vec{e}}{\partial x} \times \frac{\partial \vec{e}}{\partial y} \right) \mathrm{d}x\mathrm{d}y \,, \tag{1}$$

where $\vec{e}$ resents the unit vector of the local electric field, and $\sigma$ denotes the region that confines the skyrmions. For a skyrmion texture, the skyrmion number $N$ is an integer (±1) and can be phenomenologically understood as the number of times the vectors wrap around a unit sphere [34]. For a meron texture, the skyrmion number $N$ is a half-integer equal to ±1/2, representing the vectors can cover a half unit sphere.

### *2.2 The electric field skyrmion on the metal-air optical interface*

The two-dimensional SPPs field is an ideal physical system supporting optical skyrmions [9]. Figure 1 illustrates the configuration for SPP excitation and skyrmion generation. In this configuration, the incident structured light is a left-handed circularly polarized vortex beam (CPVB). It propagates through an objective lens with a high NA and is focused on the three-layer structure composed of a thin silver film sandwiched between air and glass substrate. The light that satisfies the wavevector matching condition will excite SPPs on the metal-air interface. It is noteworthy that the CPVB carries spin angular momentum (SAM) and orbital angular momentum (OAM), which are fundamental characteristics of electromagnetic fields [35-38].

The spin-to-orbital angular momentum conversion [38,39] occurs when the CPVB is tightly focused, which can affect the wavefront phase as well as the amplitude of near-field [40,41].

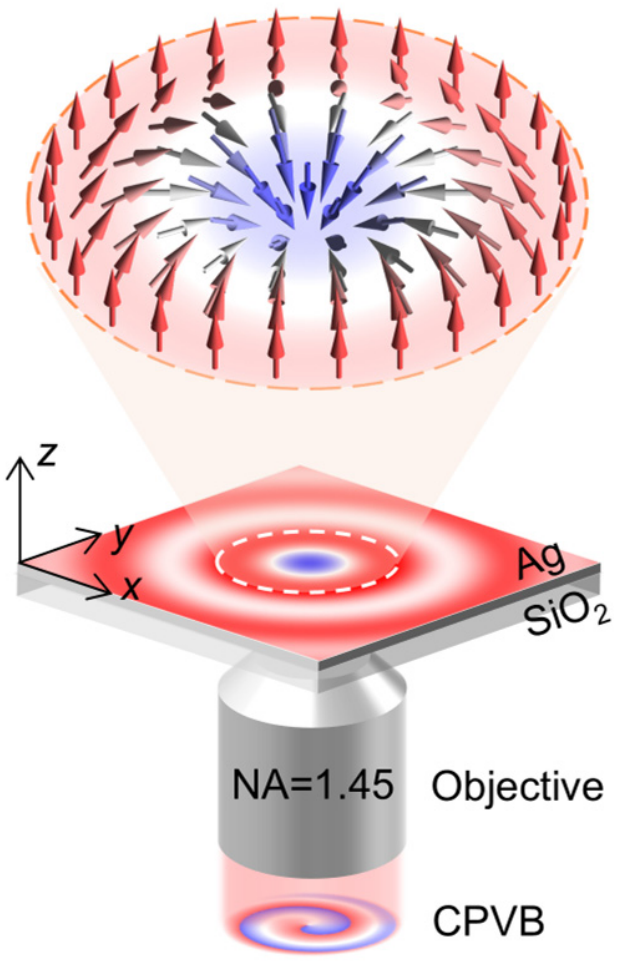

Fig. 1. Diagram for the optical generation of the electric Néel-type skyrmion.

Following the Richard-Wolf vector diffraction theory [42,43], we derived the electric field of the SPPs (see Supplement 1) excited on the metal-air optical interface, which can be written as:

$$E_\rho = \frac{\pi}{\lambda} \cdot i^m \cdot \exp(im\phi_s) \int_0^\alpha \sqrt{(cos\theta_1)} \cdot l_0(\theta_1) \cdot \exp(ik_2 cos\theta_2 (z+z_0)) \cdot \exp(-ik_1 z_0 \cos\theta_1) \\ \cdot \sin\theta_1 \cdot t_p^r(\theta_1) \cdot \cos\theta_2 \cdot [J_{m+1}(k_1 r \sin\theta_1) - J_{m-1}(k_1 r \sin\theta_1)] d\theta_1 \tag{2}$$

$$E_\varphi = \frac{\pi}{\lambda} \cdot i^{m-1} \cdot \exp(im\phi_s) \int_0^\alpha \sqrt{(cos\theta_1)} \cdot l_0(\theta_1) \cdot \exp(ik_2 \cos\theta_2 (z+z_0)) \cdot \exp(-ik_1 z_0 \cos\theta_1) \\ \cdot \sin\theta_1 \cdot t_s(\theta_1) \cdot [J_{m+1}(k_1 r \sin\theta_1) - J_{m-1}(k_1 r \sin\theta_1)] d\theta_1 \tag{3}$$

$$E_z = \frac{2\pi}{\lambda} \cdot i^{m+1} \cdot \exp(im\phi_s) \int_0^\alpha \sqrt{(\cos\theta_1)} \cdot l_0(\theta_1) \cdot \exp(ik_2 \cos\theta_2 (z+z_0)) \cdot \exp(-ik_1 z_0 \cos\theta_1) \\ \cdot \sin\theta_1 \cdot t_p^z(\theta_1) \cdot \sin\theta_2 \cdot J_m(k_1 r \sin\theta_1) d\theta_1 \tag{4}$$

Here, $m$ is the total angular momentum expressed as $m = s + l$, $\alpha$ is the maximum allowed incident angle of the objective lens, which can be calculated through $\alpha = \arcsin(\mathrm{NA}/n_i)$, $n_i$ is the refractive index of the medium. We assume that the amplitude profile of the incident beam satisfies the Bessel-Gauss distribution, which can be expressed as:

$$l_0(\theta_1) = \exp\left[-\beta_0^2 \left(\frac{\sin\theta_1}{\sin\alpha}\right)^2\right] J_1\left(2\beta_0 \frac{\sin\theta_1}{\sin\alpha}\right),$$

where $\beta_0$ is the ratio of the pupil radius and the beam waist, $J_n(x)$ denotes the $n^{th}$ order Bessel function of the first kind. The $t_p^r$, $t_s$ and $t_p^z$ are the transmission efficiencies of $E_\rho$, $E_\varphi$, and $E_z$ components through the silver film at incident angle of $\theta_1$, respectively, which can be calculated through the Fresnel transmission coefficients [44].

## 3. Results and discussion

### *3.1 The generation of the isolated electric Néel-type skyrmion*

Following the configuration in the last section, we calculated the electric field distribution from both theoretical calculation and numerical simulation. The theoretical results are calculated according to the Eqs. (2-4) and the corresponding parameters are set as follows: the SAM and OAM of the incident light are $s = 1$ and $l = -1$, the wavelength is 632.8 nm, the thickness of the silver film is 45 nm, the optical constants of silver follow the Johnson-Christy model, the refractive indices of the air and glass substrate are set to 1 and 1.515, respectively. The simulated results are calculated using the Lumerical FDTD Solutions software with the same parameters. Note that a user-designed circular mask was used for simulation to constrain the symmetry of the source since the FDTD region is rectangular cuboid, which may naturally bring in additional symmetry. The theoretical electric field distributions are shown in Figs. 2(a-c). In Figs. 2(a) and 2(c), the real part of the radial component Re($E_\rho$) and longitudinal component Re($E_z$) have circular symmetry. In Fig. 2(b), the real part of the azimuthal component Re($E_\varphi$) is almost zero, which means that the vectors will not tilt in the azimuthal direction, possessing the characteristics of Néel-type skyrmions. In addition, the phase of $E_z$ is not an intuitive spiral pattern but a binary distribution, as shown in the top right corner of Fig. 2(c). This is because the SAM and OAM of the incident light are opposite, which causes the total angular momentum $m$ to be zero. According to Eq. 4, the spiral phase term $\exp(im\phi_s)$ will also be zero, resulting in a non-spiral pattern for the phase of $E_z$. We further plotted the cross-section of the electric field components along the x-axis in Fig. 2(d) and defined the first positive peak position of Re($E_z$) as the optical skyrmion boundary, as shown by the black dotted line, which ranges from -380 nm to 380 nm. In these curves, Re($E_\rho$) is equal to zero in the center and Re($E_z$) reaches a negative peak, implying that the vector is "down" at this position. Meanwhile, Re($E_z$) shows a significant increase from the center to the boundary, indicating that the vectors will transition from a "down" state to an "up" state. The corresponding 3D vector distribution can be seen at the top of Fig. 1(a), which shows a special spin texture analogous to a hedgehog structure, namely Néel-type skyrmions. We did integration in the skyrmion region according to Eq. 1, and the skyrmion number is 0.97. In Figs. 2(e-h), we presented the simulated results, which show great agreement with the theoretical calculation and the skyrmion number is also 0.97. The theoretical and numerical skyrmion numbers are not integers which may be due to the limitations of the mesh accuracy and the nonzero azimuthal components (see Supplement 1 for detailed analysis). In general, we proved that the proposed method could generate the isolated Néel-type skyrmion and the skyrmion number is close to 1.

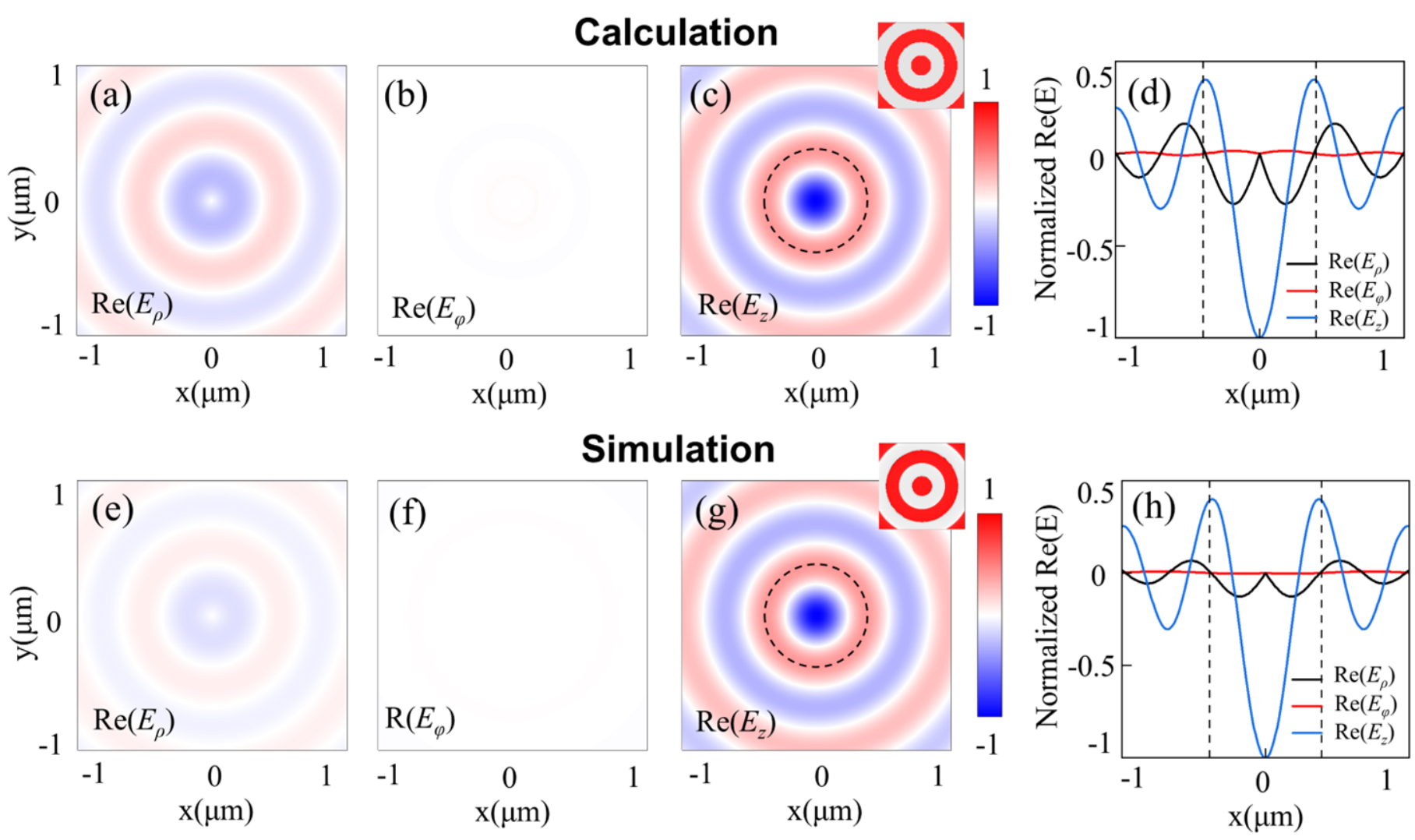

Fig. 2. (a-c) Theoretical electric field components for (a) Re($E_\rho$), (b) Re($E_\varphi$) and (c) Re($E_z$) in cylindrical coordinates, where the inset in (c) is the phase. (d) The normalized cross-section of the real part of the electric field components along the x-axis. (e-h) The corresponding electric field distributions from simulation.

### *3.2 The generation of skyrmion lattice and meron lattice constrained by symmetric apertures*

Here, we employed different apertures to constrain the symmetry of the plasmonic field to study the distribution of the electric field since the field symmetry is the key to determining the topological texture [21]. As shown in Fig. 3(a), the symmetric aperture is placed after the objective lens. The parameter $\varphi$ in the aperture plane represents the initial phase of the incident light, which is set to zero by default and can be controlled by a spatial light modulator in experiments (please refer to Supplement 1 for detailed experiment description). In the back focal plane of the objective lens, the red virtual SPP excitation ring is a bit larger than the half-side length of the square aperture, as shown in Fig. 3(b). In this condition, the originally circular SPP wave is constrained to arc-shaped SPP waves located in four corners. It propagates from corners to the center, forming the square lattice-like field after interference, as shown in the principle schematic diagram of Fig. 3(d). The simulated square lattice-like field is shown in Fig. 3(f1), where Re($E_z$) shows a distribution of alternating red and blue squares in the central region. The enlarged electric field profile and the field vectors are shown in Fig. 3(f2), where the vectors show a meron-like distribution, and the arrows point to the center in blue unit cells and point out in red unit cells. The 1D vectors along the white dashed line in Fig. 3(f2) are shown in Fig. 3(f3), which clearly visualizes the central vector variation of the meron topology. Three typical meron unit cells, identified by the black dashed square in Fig. 3(f1), were selected for the calculation of their skyrmion numbers, with the results labeled where the unit cell is located. The skyrmion number of other unit cells can be obtained by symmetry.

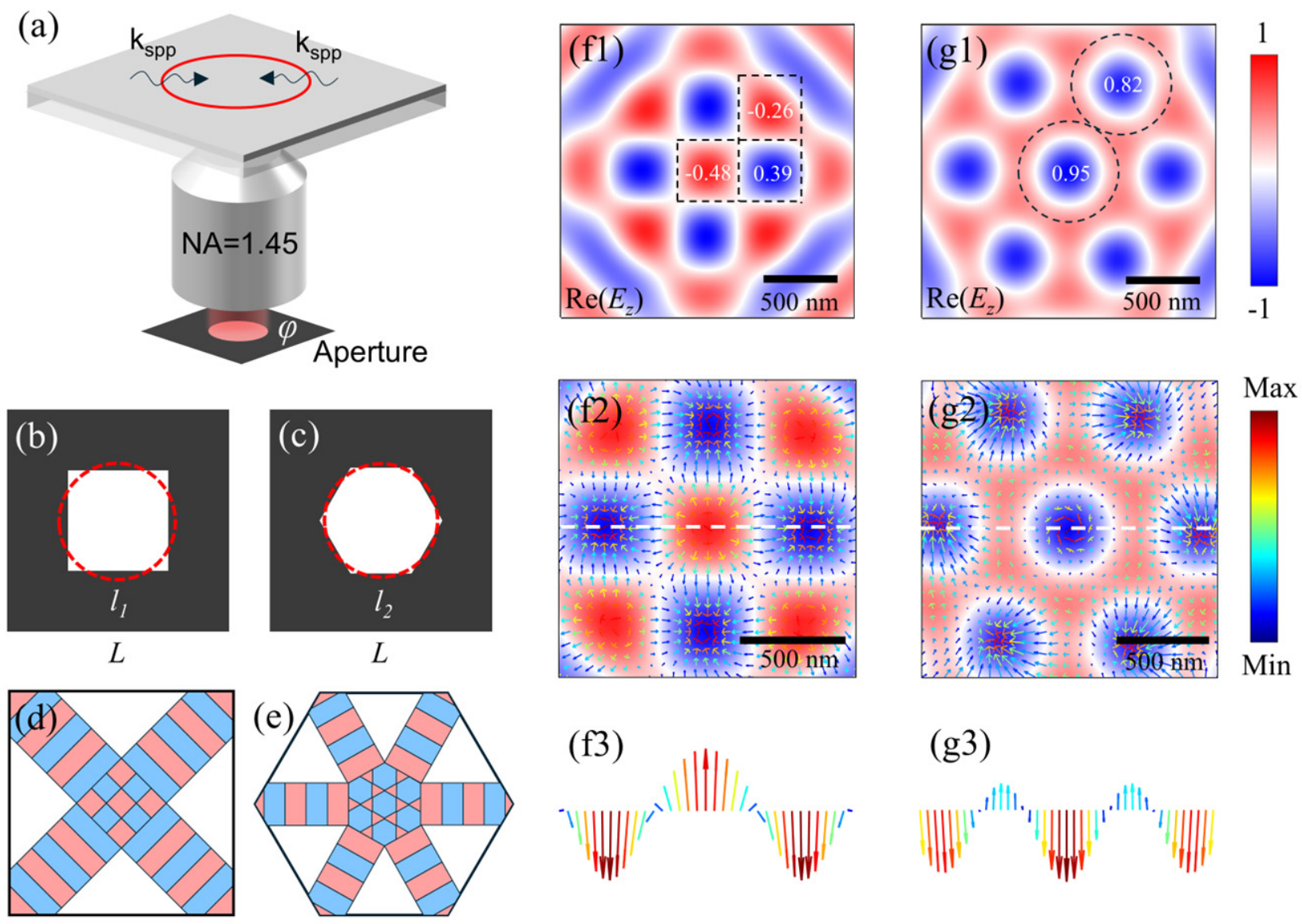

Fig. 3. (a) Optical configuration with an extra symmetric aperture. The red solid circle represents the real SPP excitation ring on the metal-air interface. The parameter $\varphi$ is the initial phase of the incident light. (b-c) The schematic diagram of reflected light at the back focal plane of the objective lens constrained by the square (b) and the hexagonal (c) apertures. The red dashed circle represents the virtual SPP excitation ring. The size parameters $l_1$, $l_2$, and $L$ are 10 μm, 6 μm, and 20 μm, respectively. (d-g) The schematic diagram of the SPP propagation and interference (d, e), the simulated distributions of Re($E_z$) (f1, g1), and the field vectors (f2-f3, g2-g3) under the symmetry constraints of the square and the hexagonal apertures, respectively. The red and blue in (d) and (e) represent Re($E_z$) greater or less than zero. The black dashed square and circle in (f1) and (g1) represent the range of meron and skyrmion unit cells, respectively. The black scale bar is 500 nm. The arrows represent the directions of the field vectors, and the color of the arrow represents the intensity of the vectors.

Similarly, we introduced a sixfold symmetry with a simple hexagonal aperture. As shown in Fig. 3(c), the virtual SPP excitation ring is a bit larger than the incircle of the hexagonal aperture, so the circular SPP wave is constrained to arc-shaped SPP waves with sixfold symmetry. It propagates to the center, forming the hexagonal lattice-like field after interference, as shown in Fig. 3(e). The simulated hexagonal lattice-like field is shown in Fig. 3(g1), where Re($E_z$) is composed of seven blue circular cells. The enlarged electric field profile and the field vectors are shown in Fig. 3(g2). The field vectors point to the center of each skyrmion-like unit cell. The 1D vectors along the white dashed line in Fig. 3(g2) are shown in Fig. 3(g3), which clearly visualizes the central vector variation of the skyrmion topology. Two typical skyrmion unit cells, identified by the black dashed circle in Fig. 3(g1), were selected for the calculation of their skyrmion numbers, with the results labeled where the unit cell is located. The skyrmion number of other unit cells can be obtained by symmetry. As a result, we generated the meron lattice and the skyrmion lattice by introducing simple square and hexagonal apertures in focused methods. The meron lattice and the skyrmion lattice are related to fourfold symmetry and sixfold symmetry of the plasmonic field, respectively.

Notably, the square aperture or the hexagonal aperture can also support the generation of the isolated skyrmion when the size of the aperture is larger than the virtual SPP excitation ring in the back focal plane of the objective lens. As shown in Fig. 4(a), the virtual SPP excitation ring in the back focal plane is inside the square aperture. In this condition, the real SPP excitation ring on the metal surface has a completely circular symmetry. The generated isolated Néel-type skyrmion is shown in Fig. 4(b), whose skyrmion number is 0.97. This situation is equivalent to the previously demonstrated skyrmion under the circular symmetry constraints in section 3.1. Defining this skyrmion as an initial state, we further investigated the influence of the initial phase of the incident light on the skyrmion topology. Keep the size of the aperture fixed. We added an extra π phase to the initial phase and the result is shown in Fig. 4(c), where Re($E_z$) shows an inversion distribution compared to that in Fig. 4(b). The field vectors point out radially and the skyrmion number is -0.97. Figure 4(d) shows the case that the virtual SPP excitation ring is smaller than the incircle of the hexagonal aperture in the back focal plane. The corresponding skyrmion topology is shown in Fig. 4(e). With an extra π phase on the initial phase, the inversed skyrmion can also be obtained, as shown in Fig. 4(f). Thus, under the symmetry constraints of the square aperture (or the hexagonal aperture), we can not only generate the meron lattice (or the skyrmion lattice) but also the isolated skyrmion using apertures with larger sizes. The topological inversion can be realized by the phase control of the incident light. These results provide the basis for subsequent transformations between different topological textures.

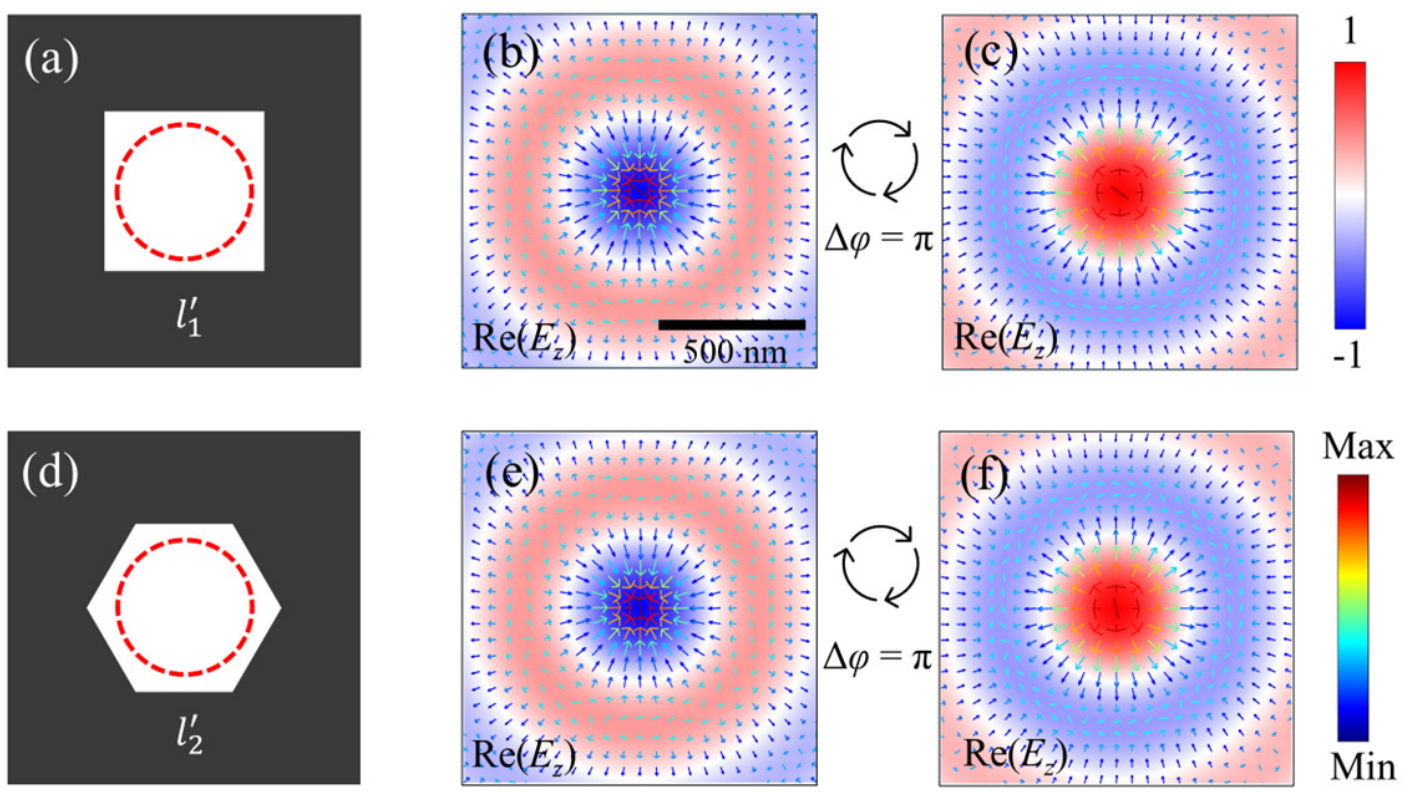

Fig. 4. (a) The schematic diagram of reflected light at the back focal plane of the objective lens constrained by the square aperture. (b-c) The isolated skyrmion under the symmetry constraints of the square aperture. The initial phases of the incident light in (b) and (c) have a π phase difference. (d-f) The corresponding results with the hexagonal aperture.

The meron lattice and the skyrmion lattice can be generated under the symmetry constraints of the simple square and hexagonal apertures. However, in that case, the two topological lattices especially the peripheral unit cells have large distortion. We studied the causes of these phenomena and found that the opening angle of the symmetric aperture plays an important role in these topologies, i.e., the topology of peripheral unit cells will distort with increasing opening angle of the symmetric aperture (please refer to Supplement 1 for more detailed quantitative analysis). The redesigned fourfold symmetry aperture is shown in Fig. 5(a), and the opening angle of the aperture is 15 degrees. The simulated electric field distributions of SPPs on the metal-air interface are shown in Figs. 5(b-d), where Re($E_x$) and Re($E_y$) show significant strip distribution, and Re($E_z$) is a mesh-like distribution, which is composed of many standard square unit cells that are uniformly distributed. The skyrmion numbers of three typical meron unit cells have been provided inside the black dashed square in Fig. 5(d), where all the values are greater than or equal to 0.45. The corresponding vector distribution is shown in Fig. 5(i), where a meron lattice is presented and composed of many red or blue square meron unit cells. We also redesigned the sixfold symmetry aperture for the optimization of skyrmion lattices. The sixfold symmetry aperture, the electric field components, and the field vectors are shown in Figs. 5(e), 5(f-h) and 5(j), respectively. The skyrmion numbers of the two typical skyrmion unit cells are both 0.97. In conclusion, we redesigned the symmetric aperture and generated the skyrmion and meron topologies with ideal topological numbers.

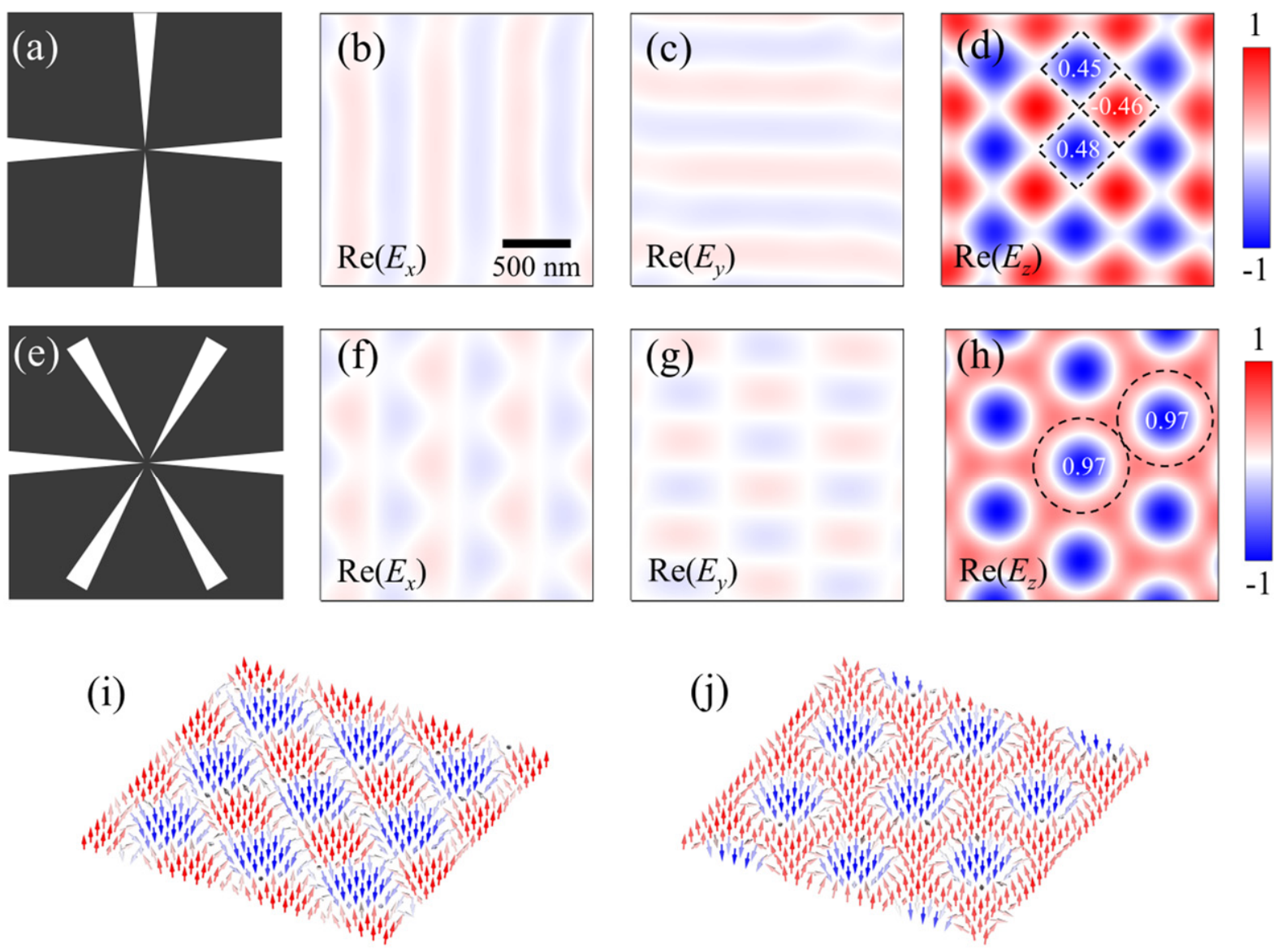

Fig. 5. (a, e) The fourfold (a) and the sixfold (e) symmetry apertures. (b-d, f-h, i-j) The normalized electric field components Re($E_x$) (b, f), Re($E_y$) (c, g), Re($E_z$) (d, h) and the 3D field vectors (i, j) under the constraints of the fourfold and the sixfold symmetry apertures. The black scale bar is 500 nm. The skyrmion number is provided inside the black dashed square and circle.

### *3.3 The phase-controlled continuous transformation constrained by well-designed symmetric apertures*

In this section, we studied the continuous transformation between different topological textures: the isolated skyrmion, the meron lattice, and the skyrmion lattice. Before this, we have demonstrated the topological inversion under the single symmetry constraint by the phase control. To realize the topological transformation, we need to provide different symmetry constraints for the plasmonic field. We first studied the transformation between the isolated skyrmion and the meron lattice. The isolated skyrmion is related to circular symmetry and the meron lattice is related to fourfold symmetry. To realize the transformation between them, the switch between circular symmetry and fourfold symmetry is necessary in our configuration. Thus, we designed a circular-fourfold symmetry aperture that is composed of an annulus superposed on the fourfold symmetry aperture, as shown in Fig. 6(a). The annulus can function as the previous circular aperture to provide circular symmetry constraints and support the generation of the isolated skyrmion. The inner and outer radii of the annulus are 4 μm and 5 μm, respectively. The size of the aperture is 20 μm. The virtual SPP excitation ring (red dashed circle) lies just on the dividing line between the two symmetries. Under the symmetry constraints of this aperture, we first generated the isolated skyrmion as shown in Fig. 6(b1). Its skyrmion number is 0.97. Here, we defined this isolated skyrmion as the initial state of the transformation, and the initial phase in this condition is zero. By gradually increasing the initial phase, we transformed this isolated skyrmion to the meron lattice in Fig. 6(b5), where all the skyrmion numbers of the typical meron unit cells are greater than or equal to 0.45. Figures 6(b2-b4) presented the continuous evolution between these two topological textures in different initial phases. The corresponding distributions of 3D vectors are shown in Figs. 6(c1-c5), which visualized the vector variation in transformation. The continuous evolution between the isolated skyrmion and the meron lattice has been shown in Visualization 1. Thus, under the symmetry constraints of the well-designed circular-fourfold symmetry aperture, we unveiled the detailed transformation process between the isolated skyrmion and the meron lattice, which intuitively reveals the relationship between the two topologies.

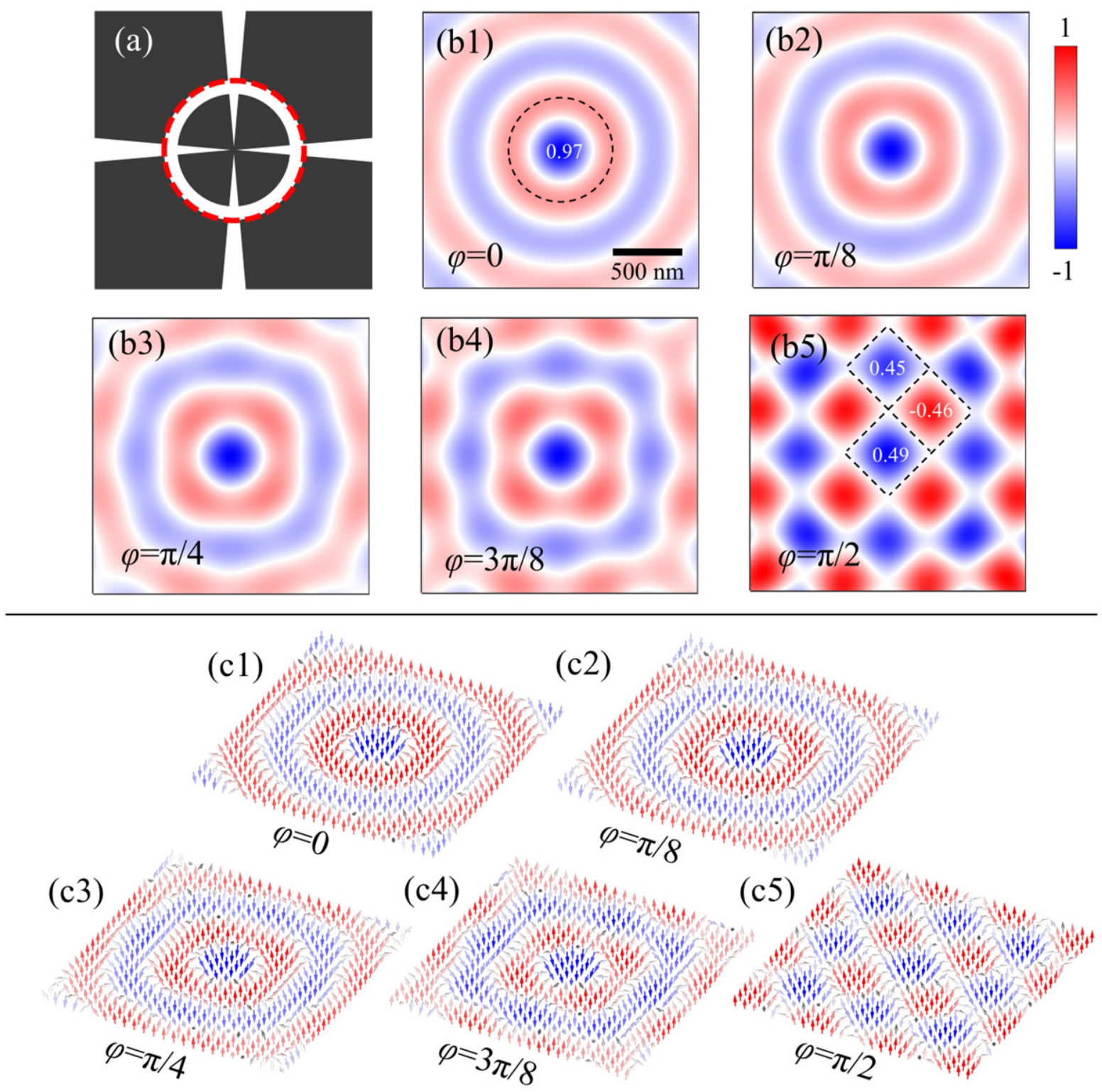

Fig. 6. The continuous transformation between the isolated skyrmion and the meron lattice. (a) The schematic diagram of reflected light at the back focal plane of the objective lens. The red dashed circle represents the virtual SPP excitation ring. The background shows the shape of the symmetric aperture for transformation. (b1-b5, c1-c5) The real part of z components of the electric field (b1-b5) and the 3D field vectors (c1-c5) under different initial phases of the incident light (see Visualization 1).

Second, we studied the transformation between the isolated skyrmion and the skyrmion lattice. These two topological textures are related to circular symmetry and sixfold symmetry, respectively. Thus, we designed a circular-sixfold aperture composed of an annulus superposed on the sixfold aperture, as shown in Fig. 7(a). The size is the same as that in Fig. 6(a), and the virtual SPP excitation ring also lies on the dividing line between the two symmetries. Under the symmetry constraints of this aperture, we simulated the electric field of SPPs on the metal-air interface. The real part of z components of the electric field is shown in Figs. 7(b1-b5), which shows the continuous transformation between the isolated skyrmion and the skyrmion lattice, i.e., from the distribution of concentric rings to the hexagonal lattice. The corresponding skyrmion numbers have been shown inside the black dashed circles, which are close to 1. The corresponding distributions of 3D vectors are shown in Figs. 7(c1-c5), which visualized the detailed vector variation in transformation. The continuous transformation between the two forms of skyrmion topology has been shown in Visualization 2. These results break the insulation of different forms of skyrmion and unveil the intrinsic relationship between skyrmion topologies.

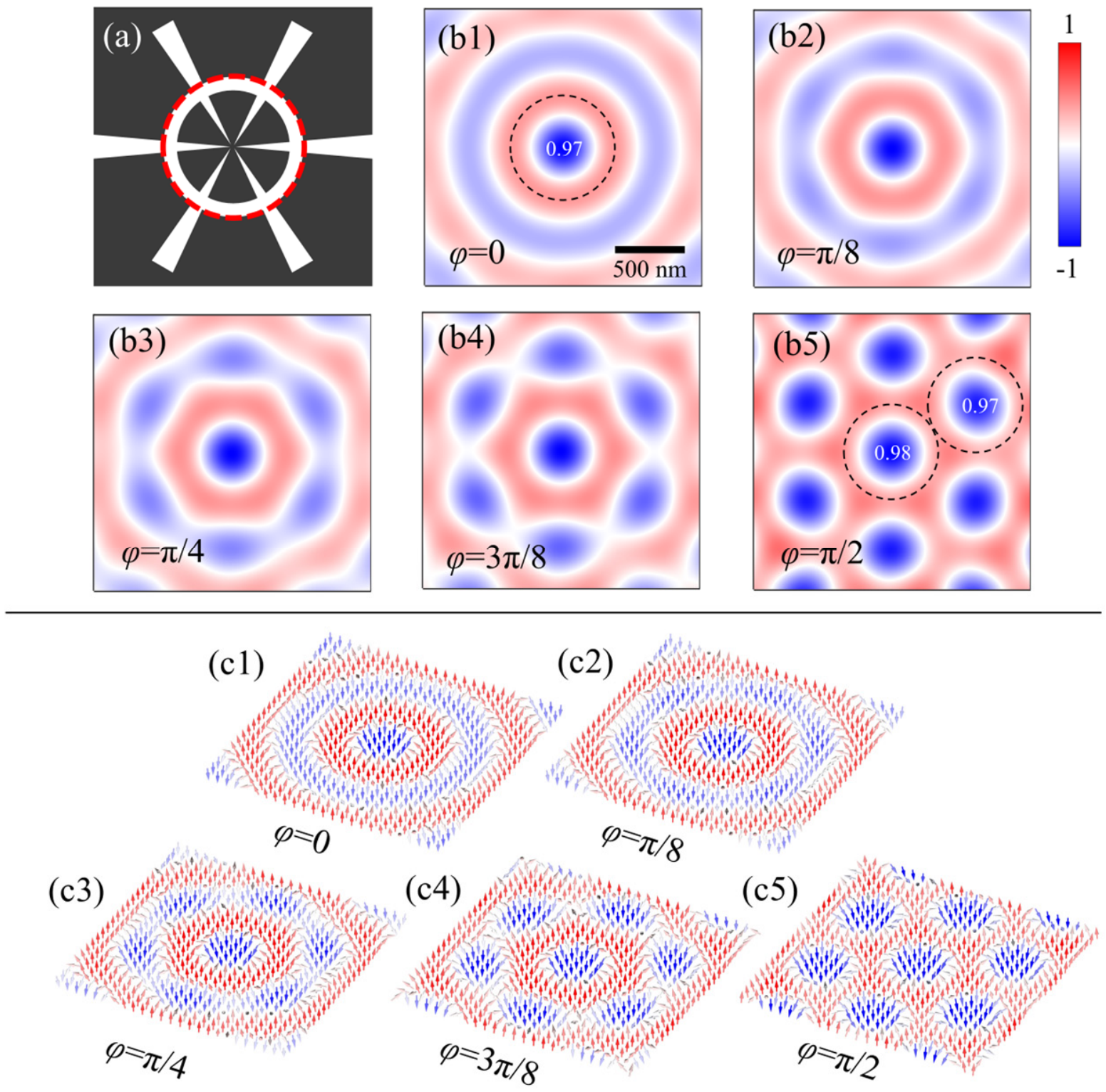

Fig. 7. The continuous transformation between the isolated skyrmion and the skyrmion lattice. (a) The schematic diagram of reflected light at the back focal plane of the objective lens. The red dashed circle represents the virtual SPP excitation ring. The background shows the shape of the symmetric aperture for transformation. (b1-b5, c1-c5) The real part of z components of the electric field (b1-b5) and the 3D field vectors (c1-c5) under different initial phases of the incident light (see Visualization 2).

Lastly, we studied the continuous transformation between the meron lattice and the skyrmion lattice. As mentioned above, the meron lattice and the skyrmion lattice are related to fourfold symmetry and sixfold symmetry, respectively. Thus, we designed a fourfold-sixfold symmetry aperture that is composed of the sixfold symmetry aperture superposed on the fourfold symmetry aperture, as shown in Fig. 8(a). This aperture has fourfold symmetry inside the range of 5 μm and sixfold symmetry outside this range. The red virtual SPP excitation ring lies on the dividing line between the two symmetries. Under the symmetry constraints of this aperture, we simulated the electric field of SPPs on the metal-air interface at different initial phases, as shown in Figs. 8(b1-b5), where the initial state is the meron lattice shown in Fig. 8(b1) and the final state is the skyrmion lattice shown in Fig. 8(b5). The skyrmion numbers of the meron unit cells and skyrmion unit cells are shown in Figs. 8(b1) and (b5), respectively. Figures 8(b2-b4) presented the intermediate states of the topological transformation. The corresponding distributions of 3D vectors are shown in Figs. 8(c1-c5), in which we can see clearly the vector variation with the change of the initial phases. The continuous transformation between the meron lattice and the skyrmion lattice has been shown in Visualization 3. In conclusion, under the symmetry constraints of the well-designed fourfold-sixfold symmetry aperture, we demonstrated the continuous transformation between the meron lattice and the skyrmion lattice by light control. These results establish and reveal the intrinsic connection between the meron lattice and the skyrmion lattice, breaking the insulation between the two topologies.

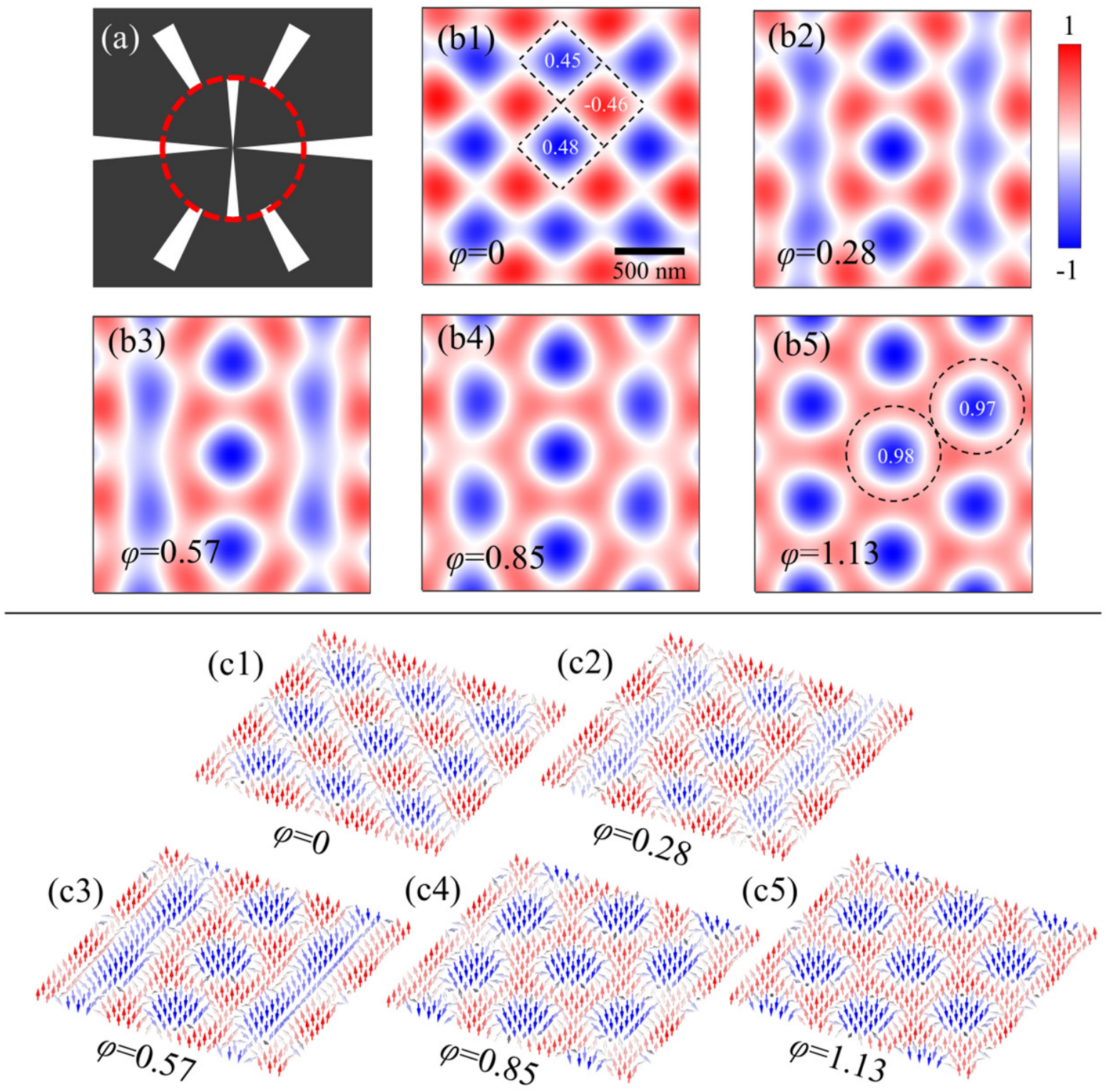

Fig. 8. The continuous transformation between the meron lattice and the skyrmion lattice. (a) The schematic diagram of reflected light at the back focal plane of the objective lens. The red dashed circle represents the virtual SPP excitation ring. The background shows the shape of the symmetric aperture for transformation. (b1-b5, c1-c5) The real part of z components of the electric field (b1-b5) and the 3D field vectors (c1-c5) under different initial phases of the incident light (see Visualization 3).

## 4. Conclusion

In this study, we realized the optical generation and continuous transformation of plasmonic field skyrmions. First, we constructed a three-layer structure that generated the isolated skyrmion with a tightly focused CPVB. We derived the analytical solutions of the electric field using the Richard-Wolf vectorial diffraction theory. Next, with the symmetry constraints of the basic square and hexagonal aperture, we generated the meron lattice and the skyrmion lattice, respectively. We further analyzed the factors that influence the distribution of skyrmion and meron topologies and provided an optimized solution for improving topological texture. After that, we designed three apertures with double symmetry, demonstrating the continuous transformation between different topological textures: the isolated skyrmion, the meron lattice, and the skyrmion lattice. Our research facilitates the comprehension of skyrmion and meron topologies, providing new insights into the symmetry of the plasmonic field, as well as showing great potential in optical information storage and encoding.

**Funding.** National Natural Science Foundation of China (62275122, 61805119); Natural Science Foundation of Jiangsu Province (BK20180469, BK20180468); Fundamental Research Funds for the Central Universities (30919011275).

**Disclosures.** The authors declare that there are no conflicts of interest related to this article.

**Data availability.** Data underlying the results presented in this paper are not publicly available at this time but may be obtained from the authors upon reasonable request.

**Supplemental document.** See Supplement 1 for support content.